\documentclass[10 pt, conference]{IEEEtran}
\IEEEoverridecommandlockouts

\usepackage{fancyhdr}

\usepackage[noadjust]{cite}

\usepackage{amsmath}
\usepackage[ruled,vlined]{algorithm2e}
\usepackage{algpseudocode}
\SetKwInput{KwInput}{Input}
\SetKwInput{KwOutput}{Output}
\usepackage{graphicx}
\usepackage{tabularx}
\usepackage{textcomp}
\usepackage{xcolor}
\usepackage{balance}
\usepackage{comment}
\def\BibTeX{{\rm B\kern-.05em{\sc i\kern-.025em b}\kern-.08em
    T\kern-.1667em\lower.7ex\hbox{E}\kern-.125emX}}
\usepackage[english]{babel}
\usepackage[utf8]{inputenc}
\usepackage{subcaption}
\usepackage{caption}
\usepackage{multirow}
\usepackage{graphics}
\usepackage{adjustbox}
\usepackage{soul}
\usepackage{enumitem}
\usepackage{tabularx}
\usepackage{array}
\usepackage{wrapfig}
\usepackage{url}
\usepackage{amsmath,amsfonts,amssymb}
\usepackage{graphicx}
\usepackage{pifont} 
\usepackage{booktabs} 

\usepackage{booktabs}
\usepackage{graphicx}
\begin{document}


\title{Weight Transformations in Bit-Sliced Crossbar Arrays for Fault Tolerant Computing-in-Memory: Design Techniques and Evaluation Framework}




\author{\IEEEauthorblockN{Akul Malhotra and Sumeet Kumar Gupta}
\IEEEauthorblockA{
\textit{Purdue University}\\
West Lafayette, Indiana, USA \\
malhot23@purdue.edu}}









\maketitle
\thispagestyle{empty}
\pagestyle{empty}

\begin{abstract}
The deployment of deep neural networks (DNNs) on compute-in-memory (CiM) accelerators offers significant energy savings and speed-up by reducing data movement during inference. However, the reliability of CiM-based systems is challenged by stuck-at faults (SAFs) in memory cells, which corrupt stored weights and lead to accuracy degradation. While closest value mapping (CVM) has been shown to partially mitigate these effects for multibit DNNs deployed on bit-sliced crossbars, its fault tolerance is often insufficient under high SAF rates or for complex tasks. In this work, we propose two training-free weight transformation techniques, sign-flip and bit-flip, that enhance SAF tolerance in multi-bit DNNs deployed on bit-sliced crossbar arrays. Sign-flip operates at the weight-column level by selecting between a weight and its negation, whereas bit-flip provides finer granularity by selectively inverting individual bit slices. Both methods expand the search space for fault-aware mappings, operate synergistically with CVM, and require no retraining or additional memory. To enable scalability, we introduce a look-up-table (LUT)-based framework that accelerates the computation of optimal transformations and supports rapid evaluation across models and fault rates. Extensive experiments on ResNet-18, ResNet-50, and ViT models with CIFAR-100 and ImageNet demonstrate that the proposed techniques recover most of the accuracy lost under SAF injection, often restoring performance to within 1–2\% of ideal baselines. Hardware analysis shows that these methods incur negligible overhead, with sign-flip leading to sub-2\% energy, latency, and area cost, and bit-flip providing higher fault resilience with modest overheads. These results establish sign-flip and bit-flip as practical and scalable SAF-mitigation strategies for CiM-based DNN accelerators.
\end{abstract}


\vspace{-0.1in}

\section{Introduction}
\label{sec-multibit:introduction}

The remarkable success of deep neural networks (DNNs) in delivering state-of-the-art performance across diverse tasks has fueled widespread deployment of DNN inference in real-world applications. However, the execution of these models on conventional von Neumann platforms (e.g., CPUs, GPUs, and TPUs) remains highly resource-intensive, incurring substantial computational, memory, and energy costs \cite{benchmarking}. These inefficiencies pose significant challenges for scaling DNN inference, particularly in scenarios constrained by power and latency requirements.

Computing-in-memory (CiM) has emerged as a promising hardware paradigm to overcome the von Neumann bottleneck that limits conventional AI accelerators. By enabling in-memory vector–matrix multiplication, the dominant kernel in DNN inference, CiM significantly reduces data transfers between compute and memory units, thereby improving both energy efficiency and latency. A wide range of CiM-based DNN accelerators have been introduced in recent years, demonstrating substantial advantages over traditional von Neumann platforms such as GPUs \cite{cim_acc1,cim_acc2,cim_llm1,cim_llm2}.

Another effective approach to reducing the cost of DNN inference is quantization of model parameters, including both weights and activations. Quantization can be incorporated directly into the training process through quantization-aware training, or applied after model convergence using post-training quantization. Importantly, CiM architectures naturally support multi-bit integer weights and activations through bit-slicing and bit-streaming, respectively, thereby enabling efficient realization of quantized models across a range of precision levels.

While CiM-based DNN accelerators offer significant advantages, their deployment must account for the impact of manufacturing defects on model performance. A notable example is stuck-at faults (SAFs), which are irreversible hard faults that corrupt memory cells within an array. Since the memory cells in CiM architectures are responsible for storing weight values, the presence of SAFs leads to incorrect weight programming and, consequently, erroneous computations that can degrade accuracy. Importantly, conventional techniques for fault tolerance developed for traditional memories may not be directly applicable in the context of CiM, where memory and computation are tightly coupled. This necessitates the development of specialized solutions to mitigate the impact of such faults in CiM-enabled systems \cite{ecc,mapping1,mapping3}.

Various solutions have been explored to mitigate this issue albeit with their own shortcomings. Several works have explored training/finetuning \cite{training2,training1} to achieve significant fault tolerance but at the expense of enormous training overheads.  Other works have pursued training-free approaches such as adding redundancy \cite{redundancy1,redundancy2}, which, although effective, lead to huge hardware costs. The works in \cite{bnnflip,retern}  utilize fault-aware weight transformations, but are mainly tailored  for low precision DNNs (such as binary \cite{bnnflip} or ternary neural networks \cite{retern}).  For high-precision (defined in this work as greater than ternary precision) DNNs,  resilience to SAFs can be achieved with Closest Value Mapping (CVM) proposed in \cite{fault-free}. CVM is a zero-overhead fault mitigation technique that maps faulty conductance values to their nearest valid representation in the quantization set. This compatibility allows the high-precision models to suppress a portion of the error introduced by SAFs without requiring retraining or additional hardware. However, despite this advantage, CVM alone is often insufficient under high fault rates and/or for more challenging workloads, as residual errors can still accumulate and cause non-trivial degradation in model accuracy.

To further improve fault resilience with minimal hardware complexity, we introduce two training-free weight transformation techniques, namely sign-flip and bit-flip. The sign-flip method selectively inverts the sign of the  weights  (stored in bit-slices in the columns of crossbar arrays), opting for the representation (original or flipped) that minimizes the SAF-induced deviation from the ideal weight values. The bit-flip method targets individual bits of the weights in crossbar columns, flipping them only when doing so yields a lower cumulative error. These techniques are designed to work synergistically with CVM, providing an additional layer of robustness without requiring retraining or extensive hardware modifications.

Furthermore, to facilitate the development and deployment of these techniques, we present a design and evaluation framework that enables rapid computation of optimal sign- and bit-flip transformations, making it practical for use in large-scale CiM-based DNN accelerators.

We evaluate the proposed techniques utilizing our framework with SAF injection experiments on 8-bit weight and activation ResNet-18, ResNet-50, and ViT models trained on both CIFAR-100 and ImageNet. While the CVM baseline experiences up to 10\% accuracy loss under severe fault conditions, our sign-flip and bit-flip methods recover most of this drop, often restoring accuracy to near-ideal (software) levels with negligible implementation overhead. The key contributions of this paper are:

\begin{itemize}
    \item We propose two novel, training-free weight transformation techniques, namely sign-flip and bit-flip, that enhance the fault tolerance of multibit DNNs deployed on bit-sliced crossbar arrays.

    \item We develop a design and evaluation framework that enables rapid simulation and optimization of weight transformations in the presence of SAFs for large-scale DNNs.

    \item We demonstrate that our proposed techniques work synergistically with CVM to significantly reduce inference accuracy loss caused by SAFs.

    \item We conduct extensive experiments on ResNet-18/50 and ViT models over CIFAR-100 and ImageNet showing that our methods recover nearly all of the accuracy lost due to SAF injection (for the fault rates considered), achieving near-ideal performance with negligible hardware overhead.
\end{itemize}

The rest of this paper is organized as follows. Section~\ref{sec-multibit:background} provides the background and discusses prior works related to SAF tolerance in CiM-based accelerators, including CVM. Section~\ref{sec-multibit:weight-transform} details the proposed weight transformations i.e. sign-flip and bit-flip, presenting the algorithm along with the hardware implications. Section~\ref{sec-multibit:results} discusses the experimental results, highlighting the effectiveness of our techniques across different models and datasets. Finally, Section~\ref{sec-multibit:conclusion} concludes the paper.

\section{Background and Related Works}
\label{sec-multibit:background}
\subsection{Deploying DNNs on bit-sliced crossbars}
\label{sec-multibit:bit-slicing}
Deploying deep neural networks (DNNs) on CiM-enabled crossbar arrays is a compelling strategy for enabling energy-efficient and high-throughput inference, particularly in resource-constrained edge settings. However, a fundamental constraint in such deployments is the limited bit precision supported by emerging memory technologies. Most DNNs operate with multi-bit weights (e.g., 8-bit), while CiM-compatible memories can typically store only a small number of bits per cell. (For instance, SRAMs are limited to binary storage. Other memories, such as ReRAMs offer the technological capabilities of storing multiple bits per device, but practically, it is challenging to compute with 8-bits per device \cite{rram_storage}.) 

To address this mismatch, a standard approach is to decompose weights using bit-slicing. If the memory supports $b$-bit storage per cell, then an $n$-bit weight must be sliced across $\lceil \frac{n}{b} \rceil$ memory cells. For instance, an 8-bit weight stored in a 2-bit ReRAM technology would require four separate slices, each storing 2 bits. This general form of bit-slicing enables flexible deployment of DNNs across memory technologies with varying precision capabilities.

Among these, a common and widely supported configuration is binary bit-slicing, where each memory cell stores just 1 bit \cite{hcim,vesti}. This configuration is required for SRAM, which inherently supports only 1 bit per cell. It is also commonly used in multi-bit memories such as ReRAM, as operating in binary mode offers larger noise margins and improved state distinguishability \cite{unbalanced}, especially in the context of CiM. In this approach, an $n$-bit weight is split across $n$ crossbar arrays, each storing one binary weight plane $\mathbf{W}^{(k)}$ corresponding to bit position $k$ with significance $2^k$.

Activations, which are applied to CiM arrays as analog voltages, also face similar precision limitations. In theory, multi-level voltage encoding can represent $m$-bit activations using finely spaced voltage levels (e.g., $0$, $\frac{1}{2^m-1}V_{\mathrm{DD}}$, $\frac{2}{2^m-1}V_{\mathrm{DD}}$, ... $V_{\mathrm{DD}}$). However, such schemes suffer from reduced voltage margins and increased susceptibility to noise and sensing errors. These challenges are further amplified in crossbar designs in which the voltages are applied at the gate of the transistors in the bit-cell (rather than the current-carrying terminals), to achieve higher CiM robustness \cite{bitcells}. Such designs require non-linearly spaced activation voltages to compensate for the inherent non-linearity of the transistor, thereby complicating digital-to-analog converter (DAC) design \cite{twin8t}. Increasing the supply voltage $V_{\mathrm{DD}}$ to compensate can mitigate sensing errors but leads to higher power consumption. 

Thus, in this paper, we focus on binary bit-slicing for weights and binary bit-streaming for activations for deploying multibit DNNs. In binary bit-streaming, each $m$-bit activation is decomposed into $m$ sequential binary values $\mathbf{a}^{(l)}$, where $l \in \{0, 1, \dots, m-1\}$ denotes the bit index. Each bit-plane $\mathbf{a}^{(l)}$ is applied sequentially using two voltage levels: 0 and $V_{\mathrm{DD}}$.
 This averts the use of DACs, significantly reducing power and design complexity while improving signal robustness.

Each sub-computation involves computing the partial dot-product $\mathbf{W}^{(k)} \cdot \mathbf{a}^{(l)}$ between one activation bit vector $\mathbf{a}^{(l)}$ and one weight bit-plane $\mathbf{W}^{(k)}$. The complete dot-product is then reconstructed by aggregating these partial results using near-memory shift-and-add logic. When both weights and activations are unsigned, the reconstruction follows:

\begin{equation}
y = \sum_{k=0}^{n-1} \sum_{l=0}^{m-1} 2^{k+l} \left( \mathbf{W}^{(k)} \cdot \mathbf{a}^{(l)} \right)
\label{eq:multibit-1}
\end{equation}

However, in practice, weights and activations are often signed. In some cases, two separate arrays (positive and negative arrays) are used to store the negative and positive weights \cite{fault-free,redundancy1}, and the two dot products computed from them are subtracted to obtain the overall dot product. In this case, the weights can be stored in the unsigned format, and Eq.~\ref{eq:multibit-1} is valid. However, this approach effectively doubles the memory footprint, requiring two arrays for each weight vector, which is inefficient in terms of area and energy. A more compact and hardware-efficient method for representing signed weights is to use two’s complement encoding. In this format, the most significant bit (MSB) represents a negative contribution (e.g., $-2^{n-1}$ for an $n$-bit number), which must be explicitly incorporated into the accumulation logic. To correctly reconstruct the dot-product using two’s complement operands, the partial sums involving MSBs of the weights and activations must be scaled negatively. The resulting expression is:

\begin{align}
y =\ &\sum_{k=0}^{n-2} \sum_{l=0}^{m-2} 2^{k+l} \left( \mathbf{W}^{(k)} \cdot \mathbf{a}^{(l)} \right) \nonumber \\
& -\sum_{l=0}^{m-2} \left(2^{n-1 + l} \right) \left( \mathbf{W}^{(n-1)} \cdot \mathbf{a}^{(l)} \right) \nonumber \\
&- \sum_{k=0}^{n-2} \left(2^{k + m-1} \right) \left( \mathbf{W}^{(k)} \cdot \mathbf{a}^{(m-1)} \right) \nonumber \\
&+ \left(2^{n-1 + m-1} \right) \left( \mathbf{W}^{(n-1)} \cdot \mathbf{a}^{(m-1)} \right)
\end{align}

Here, $\mathbf{W}^{(n-1)}$ denotes the MSB bit-plane of the weight (the sign bit), and $\mathbf{a}^{(m-1)}$ denotes the MSB bit-stream of the activation. The first term captures interactions between all lower bits, while the remaining terms account for the negative contributions from the MSBs of weights and/or activations. In DNNs where activations are guaranteed to be non-negative (e.g., ReLU-activated networks), unsigned representation can be used for activations, and the equation simplifies accordingly. In this paper, we utilize the two's complement encoding for signed weights as well as for signed activations (where applicable).

\subsection{Stuck-at faults (SAFs) in CiM-based DNN accelerators}
\label{sec:stuck_at_faults}

With continued technology scaling and the emergence of novel memory architectures, the characterization and mitigation of stuck-at faults (SAFs) have become increasingly important. Conventional strategies for fault tolerance are often ineffective in CiM-enabled systems, where memory and computation are tightly intertwined, thereby motivating the need for CiM-specific SAF mitigation techniques. SAFs arise when a binary memory cell becomes permanently fixed to a logic ‘0’ (High-Resistance State, HRS) or logic ‘1’ (Low-Resistance State, LRS), referred to as stuck-at-0 (SA0) and stuck-at-1 (SA1) faults, respectively. Depending on the relationship between the fault state and the intended data, SAFs can be categorized as either masked or unmasked. Unmasked faults occur when the fault state conflicts with the desired stored value (e.g., an SA1 fault in a cell intended to store ‘0’), resulting in functional errors, whereas masked faults occur when the fault state matches the stored value and hence remain benign. Therefore, only unmasked faults directly contribute to computation errors in CiM-enabled memory arrays.

Several approaches have been proposed to mitigate the impact of SAFs in DNN accelerators. One class of methods enhances fault tolerance through fault-aware training or fine-tuning of the network. For instance, the works in \cite{tosa,training2} employ retraining of DNN weights under fault injection to restore accuracy, while \cite{training1} introduces drop-connect during training to improve the network’s robustness to SAFs. Although these strategies demonstrate strong effectiveness, they incur the high computational cost associated with DNN training and rely on access to labeled training data, which may not always be available in practical scenarios.

In contrast, approaches that do not rely on retraining (hereafter referred to as training-free approaches) circumvent this limitation. A common training-free approach involves introducing artificial redundancy, wherein additional rows or columns are allocated to correct or compensate for SAF-induced errors \cite{redundancy1,redundancy2,extra_row}. However, redundancy-based techniques often incur significant overheads; for instance, the energy cost in \cite{redundancy1} ranges from approximately $24\%$ to $112\%$. To address this, the work in \cite{redundancy3} proposes an overhead-free scheme that employs structured pruning to reclaim storage resources, which are subsequently repurposed for adding redundant rows and columns. Nevertheless, this approach still requires model retraining to recover the accuracy degradation caused by pruning.

Another training-free strategy to tackle SAFs is to utilize fault-aware weight mapping. The weights are mapped onto the CiM macros in a way that minimizes the number of unmasked SAFs. The works in \cite{mapping1,fault-free,mapping3} show improved accuracies for inference with SAFs.

Closest Value Mapping (CVM) is a CiM-compatible fault-aware weight mapping technique that applies to both dual-array and two’s complement weight storage formats, offering broad compatibility \cite{fault-free,faq}. Notably, it achieves improved SAF tolerance without incurring any hardware overhead. (Note, the chip-specific SAF maps need not be stored in on-chip memory, as the SAF information is used only during weight deployment). In the following subsection, we present the details of this technique.

\subsection{Baseline SAF tolerance enhancement: Closest value mapping}
\label{sec-multibit:closestvalue}

Closest Value Mapping (CVM) is a training-free, software-based technique designed to improve the fault tolerance of CiM accelerators in the presence of stuck-at faults (SAFs) at no hardware overhead \cite{fault-free}.

Given a set of $M$ ideal weights $W_{\mathrm{target}} = (w_1, w_2, \dots, w_M)$ of $n$-bit precision, the objective of CVM is to find, for each $w_i$, the closest legal value $w_i^{\mathrm{mapped}}$ that can be reliably programmed into a faulty memory cell. The legality of a candidate weight value depends on the SAFs present in each bit position, described by a stuck-at mask $S \in {-1, 0, 1}^{M \times n}$, where $-1$ denotes a stuck-at-0 fault, $1$ a stuck-at-1 fault, and $0$ indicates a fault-free bit.  The SAF information
can be obtained in advance using standard fault diagnosis procedures commonly employed during chip testing (and may be stored in an off-chip memory and retrieved during weight deployment on the crossbar arrays) \cite{fault_testing}. 

The CVM algorithm, as described in Algorithm~\ref{algo:cvm}, begins by enumerating all $2^n$ possible $n$-bit weight values and replicating them across $M$ rows to create a candidate matrix of shape $M \times 2^n$. Each row corresponds to one weight location and contains all possible weight values. A legality check is then performed to determine which candidate weights are valid considering the SAF pattern in $S$, producing a Boolean legality mask. The absolute difference between each target (ideal) weight and each candidate is then computed to form an error matrix. Illegal candidates are penalized by assigning infinite error. Finally, for each row, the candidate with the minimum error is selected as the final mapped value.

\begin{algorithm}[t]
\caption{Closest Value Mapping (CVM)}
\SetKwInput{KwInput}{Inputs}
\SetKwInput{KwResult}{Outputs}
\SetKw{KwRet}{Return} 
\label{algo:cvm}

\tcc{\textbf{Inputs:} $W_{\mathrm{target}}$ - $M$ size vector of ideal $n$-bit weights;
     $S$ - $M{\times}n$ stuck-at mask (1 = SA1, $-1$ = SA0, 0 = fault-free).\\ 
     \textbf{Output:} $W_{\mathrm{mapped}}$ - weights after closest-value mapping.}

\KwInput{$W_{\mathrm{target}}\!=\!(w_1,\dots,w_M)$}

\KwInput{$S\!\in\!\{-1,0,1\}^{M\times n}$}

\KwResult{$W_{\mathrm{mapped}}\in\mathbb{Z}^{M}$}

\textbf{Step 1 - Enumerate codes}

$\textit{cand} \gets (0,1,\dots,2^{n}-1)$              \tcp*{all $n$-bit numbers}

$CAND \gets \textsc{TileRow}(\textit{cand},\,M)$       \tcp*{$M\times2^{n}$ candidate matrix}

\textbf{Step 2 - Legality test}

$Legal \gets \textsc{IsLegal}(CAND,\,S)$               \tcp*{Boolean mask}

\textbf{Step 3 - Error computation}

$Ideal \gets \textsc{RepeatCol}(W_{\mathrm{target}},\,2^{n})$ \tcp*{broadcast $W_{\mathrm{target}}$}

$Err \gets \lvert CAND - Ideal\rvert$                       \tcp*{absolute error}

$Err \gets Err + (1-Legal)\cdot\infty$                      \tcp*{penalise illegal codes}

\textbf{Step 4 - Select closest value }

$idx \gets \operatorname*{arg\,min}_{c}\,Err$          \tcp*{min-error index per row}

$W_{\mathrm{mapped}} \gets \textsc{RowSelect}(CAND,\,idx)$ \tcp*{gather winners}

\KwRet $W_{\mathrm{mapped}}$\; \DontPrintSemicolon

\tcc*[l]{\textbf{Helper ops:} 
      \textsc{TileRow}$(v,M)$ - replicate row vector $v$ $M$ times;
      \textsc{RepeatCol}$(x,N)$ - broadcast column vector $x$ across $N$ columns;
      \textsc{RowSelect}$(A,idx)$ - for each row $i$, return $A[i,\,idx_i]$.}

\end{algorithm}

CVM is highly parallelizable, as all steps (performed in software), viz. enumeration, legality testing, error computation, and selection can be expressed as vectorized matrix operations and executed efficiently on parallel compute platforms such as GPUs. The primary memory requirement is the candidate matrix, which has size $M \times 2^n$. While this can become large if $M$ represents all weights in the DNN or for large $n$, the technique remains practical in most real-world scenarios. In quantized DNNs, weight precision can be typically limited to $n \leq 8$ while maintaining near full-precision accuracy, making the candidate space tractable. Moreover, applying CVM on a layer-by-layer basis keeps $M$ in check, enabling scalable deployment even for large models.

\begin{figure}[t!]
\centering
  \includegraphics[width = 0.9\linewidth]{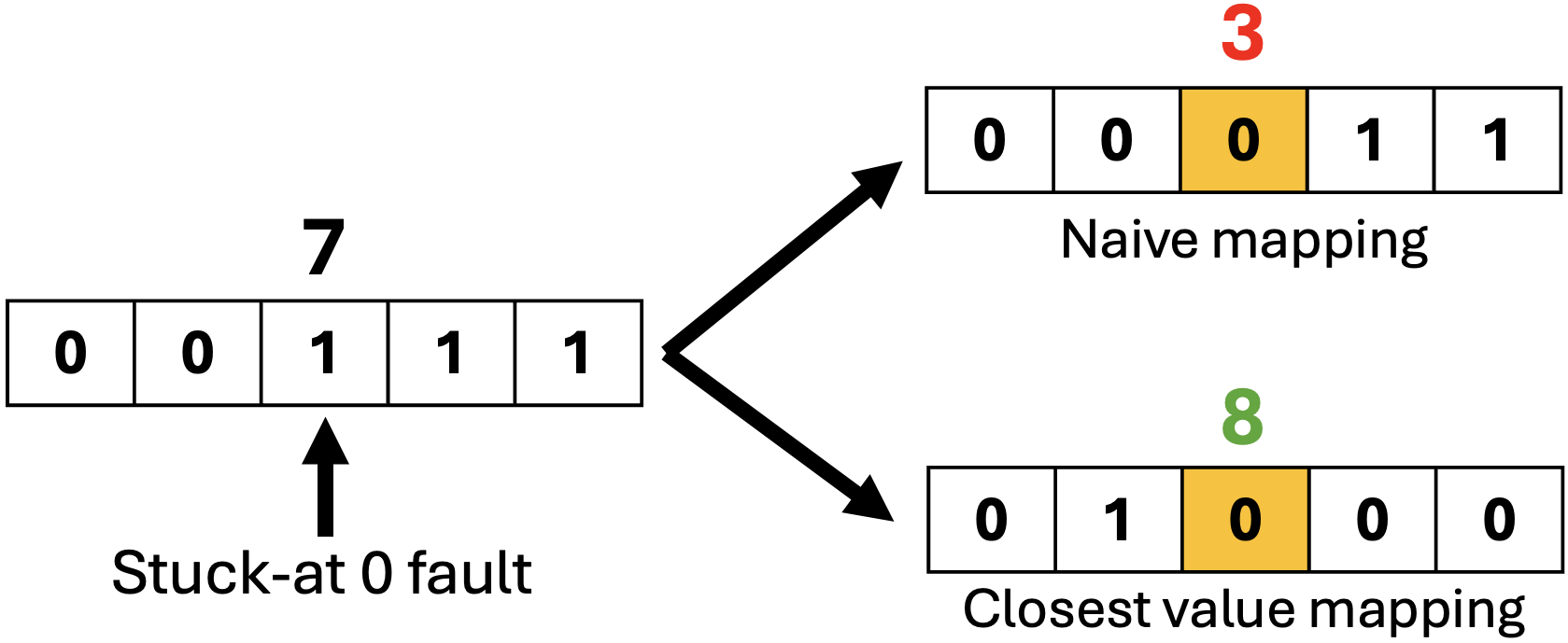}
  \caption{An example illustrating closest value mapping (CVM).}
  \label{fig-multibit:cvm}
\end{figure}

In case the accelerator is afflicted by SAFs, some of the weights will not be written correctly into the memory arrays. Let us call the ideal weight we want to write as $w_{ideal}$, and the weight that is written into the memory array (impacted by SAFs) as $w_{hardware}$. In closest value mapping (CVM), we determine the closest weight value to $w_{ideal}$ that we can write into the memory, thus minimizing error $\lvert w_{ideal} - w_{hardware} \rvert$. This is done for each weight and corresponding fault pattern in the DNN. Figure~\ref{fig-multibit:cvm} describes an example of CVM. It can be observed that that the intended weight value is 7, and with the stuck-at 0 fault at that location, a naive mapping would lead to 3 being stored. With CVM, we identify that 8 is the closest value that can be stored given the fault pattern. Thus, CVM reduces the SAF-induced error when the DNN weights are mapped onto the accelerator.

In this paper, we adopt CVM as the baseline fault-aware mapping strategy and investigate how its effectiveness can be further enhanced through two weight transformation techniques: sign-flip and bit-flip. These transformations operate in conjunction with CVM to significantly reduce SAF-induced errors in DNNs deployed on bit-sliced crossbar arrays, all without requiring any retraining. Note that CVM does not apply to binary- or ternary-precision models. This is because those models have only two or three legal codewords per weight, leaving no meaningful “nearest intact value” to choose. In contrast, higher precision weights provide $2^n$ codewords, enabling CVM to pick, for every weight and SAF pattern, the stored value whose decoded magnitude deviates least from the ideal. 

\section{Proposed Weight Transformations}
\label{sec-multibit:weight-transform}
\subsection{Sign-Flip}
\label{sec-multibit:signflip}

\begin{algorithm}[t]
\caption{The Sign-Flip Algorithm}
\SetKwInput{KwInput}{Inputs}
\SetKwInput{KwResult}{Outputs}
\label{algo:signflip}

\tcc{\textbf{Inputs:} $W_{\mathrm{target}}\!\in\!\mathbb{Z}^{M\times K}$ - ideal $n$-bit weights of a layer;\\
      \hspace{9mm}$S\!\in\!\{-1,0,1\}^{M\times K \times n}$ - SAF mask;\\
      \hspace{9mm}$row\_len$ - rows per memory array.\\
      \textbf{Outputs:} $W_{\mathrm{mapped}}\!\in\!\mathbb{Z}^{M\times K}$ - mapped weights;\\
      \hspace{10mm}$col\_flip\!\in\!\{0,1\}^{\lceil M / row\_len\rceil \times K}$ - column-flip mask.}

\BlankLine
$\mathbf{w}_{\mathrm{flat}}\leftarrow\textsc{Flatten}(W_{\mathrm{target}})$\;
$\mathbf{W}_{\pm}\leftarrow
\begin{bmatrix}
  \mathbf{w}_{\mathrm{flat}}\\[2pt]
 -\mathbf{w}_{\mathrm{flat}}
\end{bmatrix}$\tcp*{($2\times M \times K)$}

$\mathbf{S}_{\pm}\leftarrow
      \textsc{TileRow}\bigl(\textsc{Flatten}(S),\,2\bigr)$\;

$\mathbf{W}_{\mathrm{cvm}}\leftarrow
      \textsc{CVM}\!\bigl(\mathbf{W}_{\pm},\mathbf{S}_{\pm}\bigr)$\;

$\bigl[\,W_{+},\,W_{-}\bigr]\leftarrow
      \textsc{Reshape}\!\bigl(\mathbf{W}_{\mathrm{cvm}},\,2,M,K\bigr)$\;

$W_{\mathrm{mapped}}\leftarrow\mathbf{0}$\;

\For{$c\leftarrow 0$ \KwTo $\bigl\lceil M / row\_len\bigr\rceil - 1$}{
    $R\leftarrow c\cdot row\_len:\min\!\bigl((c{+}1)\cdot row\_len,M\bigr)-1$\tcp*{row slice}

    $E_{+}\leftarrow \sum^{R}\!\bigl(\lvert W_{+}[R,:]-W_{\mathrm{target}}[R,:]\rvert\bigr)$\;
    $E_{-}\leftarrow \sum^{R}\!\bigl(\lvert W_{-}[R,:]- (-W_{\mathrm{target}}[R,:])\rvert\bigr)$\;

    $col\_flip[c,:]\leftarrow(E_{-}<E_{+})$\tcp*{$1\times K$ Boolean} 

    $W_{\mathrm{mapped}}[R,:]\leftarrow
        (1-col\_flip[c,:])\odot W_{+}[R,:]\;+\;col\_flip[c,:]\odot W_{-}[R,:]$\;

    
}

\KwRet $W_{\mathrm{mapped}},\;col\_flip$\;
\DontPrintSemicolon
\tcc*[l]{\textbf{Helper ops:}
  \textsc{Flatten} - row-major vectorisation; 
  \textsc{TileRow}$(v,M)$ - replicate row vector $v$ $M$ times; 
  \textsc{Reshape}$(\cdot,2,M,K)$ - $\rightarrow(2,M,K)$ then split; 
  $\odot$ - element-wise product}
\end{algorithm}

The sign-flip technique rests on a straightforward identity: for any input-current vector~$I$ and weight column~$W$,
\[
\sum I\cdot W \;=\; -\!\sum I\cdot(-W).
\]
Because a dot product obtained with $-W$ can be corrected simply by multiplying it by $-1$ in the digital accumulator, every column in the weight matrix may be stored as either $W$ or $-W$ without affecting functional correctness, as long as appropriate post-processing is applied. Having both the negative and positive representation available introduces a new degree of freedom in how weights are mapped onto faulty memory arrays. Thus, while CVM selects a weight value from among a set of candidate values to minimize SAF-induced error, sign-flip adds another dimension to this, effectively doubling the candidate space. In other words, the proposed sign-flip provides another option (i.e. the negated counterpart in addition to the original weight) for better alignment of the weight mapping with the SAF pattern.

The sign-flip algorithm, shown in Algorithm~\ref{algo:signflip}, consolidates this expanded search. It first constructs a $2 \times M \times K$ candidate tensor by stacking $W_{\text{target}}$ and $-W_{\text{target}}$, and passes it to the CVM subroutine to obtain the closest-value mapped versions, denoted by $W_{+}$ and $W_{-}$. These represent the optimal mapped forms of the original and negated weights, respectively, under the given SAF configuration.

Next, the sign-flip determines whether to deploy weights from $W_{+}$ or $W_{-}$ on each weight column in the hardware. For that, it partitions $W_{+}$ and $W_{-}$ into chunks of size $row\_len \times K$, where $row\_len$ is the number of rows in the memory sub-arrays. Since the dot products computed by each sub-array column are independent of each other, all the weights columns can be processed in parallel. For each weight column in a chunk, the algorithm computes the total reconstruction error for both candidates: $E_+$ is the sum of absolute differences between $W_{+}$ and $W_{\text{target}}$, and $E_-$ is the corresponding error for $W_{-}$. The version with the lower error is selected and copied into the final mapped weight matrix $W_{\text{mapped}}$.

Additionally, a binary mask $col\_flip$ of size $\lceil M / row\_len \rceil \times K$ (one bit per weight column) is generated to record the polarity decision for each weight column in every array. A value of 0 indicates that the original weights were used, while a value of 1 indicates that the negated weights were selected and the corresponding CiM output must be flipped post-accumulation.

In summary, sign-flip augments the solution space by introducing polarity as an additional axis of optimization, thereby facilitating the discovery of more robust weight mappings under SAF constraints. This improvement comes with minimal hardware overhead: it requires storing $col\_flip$ for each memory array in a peripheral register and incorporating a simple two’s complement unit to negate the CiM output when necessary. These hardware implications are discussed in detail in Section~\ref{sec-multibit:hardware-imp}.

\subsection{Bit-Flip}
\label{sec-multibit:bitflip}

\begin{algorithm}[t]
\caption{The Bit-Flip Algorithm}
\SetKwInput{KwInput}{Inputs}
\SetKwInput{KwResult}{Outputs}
\label{algo:bitflip_1shot}

\tcc{\textbf{Inputs:} $W_{\mathrm{target}}\in\mathbb{Z}^{M\times K}$ - ideal $n$-bit weights\;
      $S\in\{-1,0,1\}^{M\times K\times n}$ - SAF mask\;
      $row\_len$ - rows per memory array.\\
      \textbf{Outputs:} $W_{\mathrm{mapped}}\in\mathbb{Z}^{M\times K}$ - mapped weights\;
      $b\_{\mathrm{flip}}\in\{0,1\}^{\,n\times\lceil M/row\_len\rceil\times K}$ - bit-flip mask.}

\BlankLine
$\mathbf{w}_{\mathrm{flat}}\leftarrow\textsc{Flatten}(W_{\mathrm{target}})$\;

\For{$j\leftarrow 0$ \KwTo $2^{n}-1$}{
    $\mathbf{W}_{\mathrm{cand}}[j,:,:]\leftarrow
        \mathbf{w}_{\mathrm{flat}}\;\oplus\;j$ \tcp*{vector-XOR with mask $j$}
}

$\mathbf{S}_{\mathrm{cand}}\leftarrow
      \textsc{TileRow}(\textsc{Flatten}(S),\,2^{n})$\;
$\mathbf{W}_{\mathrm{cvm}}\leftarrow
      \textsc{CVM}(\mathbf{W}_{\mathrm{cand}},\,\mathbf{S}_{\mathrm{cand}})$\tcp*{$[2^{n} \times M \times K]$}
$\mathbf{W}_{\mathrm{cvm}}\leftarrow
      \textsc{Reshape}(\mathbf{W}_{\mathrm{cvm}},\,2^{n},M,K)$\;

$W_{\mathrm{mapped}}\leftarrow\mathbf0$\;
$b\_{\mathrm{flip}}\leftarrow\mathbf0$\;

\For{$c\leftarrow 0$ \KwTo $\lceil M/row\_len\rceil-1$}{
    $R\leftarrow c\cdot row\_len : \min\!\bigl((c{+}1)row\_len,\,M\bigr)-1$\tcp*{row slice}

    $\mathbf{E}\leftarrow
      \sum_{R}\lvert
        \mathbf{W}_{\mathrm{cvm}}[:,R,:] - W_{\mathrm{target}}[R,:]
      \rvert$ \tcp*{$2^{n}\times K$}

    $\mathbf{idx}\leftarrow
      \operatorname*{arg\,min}_{j}(\mathbf{E}[j,:])$\;

    $W_{\mathrm{mapped}}[R,:]\;\leftarrow\;
        \textsc{Gather}\bigl(\mathbf{W}_{\mathrm{cvm}}[:,R,:],\;\mathbf{idx}\bigr)$\;

    $\mathbf{b}\;\leftarrow\;(0,1,\dots ,n{-}1)^{\mathsf T}$
    
    $b\_{\mathrm{flip}}[:,c,:]\;\leftarrow\;
        \bigl\lfloor\mathbf{idx}/2^{\mathbf{b}}\bigr\rfloor\bmod2$\tcp*{$[n\times\lceil M/row\_len\rceil\times K]$}     
    }

\KwRet $W_{\mathrm{mapped}},\,b_{\mathrm{flip}}$\;
\DontPrintSemicolon
\tcc*[l]{\textbf{Helper ops (same as sign-flip):}
  \textsc{Flatten}, \textsc{TileRow}, \textsc{Reshape};\;
  $\oplus$ - element-wise XOR. \\
  $\lvert\cdot\rvert$ - absolute value.\\
  \textsc{Gather}$(T,i)$ - pick rows/columns of tensor $T$ along the first axis using index vector $i$.}
\end{algorithm}

The bit-flip technique builds on the core insight that each individual bit slice of a multi-bit weight can be selectively flipped to better align with the SAFs characteristics of a memory array,without altering the correctness of the final dot product. Unlike sign-flip, which applies a polarity decision to all the bits of the weight, bit-flip operates at the granularity of an individual bit slice, dramatically increasing the degrees of freedom available during weight mapping. It relies on the identity:
\[
\sum I \cdot \left(1 - W^{(j)}\right) = \sum I - \sum I \cdot W^{(j)},
\]
where $W^{(j)} \in \{0,1\}^{M \times K}$ denotes the $j$-th binary slice of the weight column and the sum is taken across the input vector $I$, which is common to the entire weight matrix. Because the contribution of a flipped slice can be recovered by subtracting its dot product from the input sum $\sum I$, each bit slice can be stored in either its original or complemented form without affecting the functional correctness of the final dot product. This introduces per-slice polarity control-across all $n$ slices of each weight, enabling an exponential increase in the search space to $2^n$ possible representations per weight. In contrast to the binary flexibility of sign-flip, this fine-grained control significantly enhances the potential for SAF-tolerant mappings.

The bit-flip algorithm, outlined in Algorithm~\ref{algo:bitflip_1shot}, systematically explores this expanded solution space in three stages. First, it flattens the ideal weight matrix $W_{\mathrm{target}}$ and XORs it with every integer mask $j \in \{0, 1, \dotsc, 2^n - 1\}$ to construct a candidate tensor of size $2^n \times M \times K$, enumerating all possible patterns of slice flips. Second, this tensor, along with a tiled SAF mask, is passed to the CVM subroutine, which returns $2^n$ closest-value-mapped tensors $\mathbf{W}_{\mathrm{cvm}}[j,:,:]$. Third, for each memory sub-array (a row block of height $row\_len$), the algorithm computes in parallel the reconstruction error for each candidate across all $K$ weight columns as
\[
E_{j,k} \;=\; \sum_{r \in R}
              \lvert
                \mathbf{W}_{\mathrm{cvm}}[j,r,k] -
                W_{\mathrm{target}}[r,k]
              \rvert
\]
The algorithm selects the index $j$ that minimizes $E_{j,k}$, and copies the corresponding weights into the final mapped matrix $W_{\mathrm{mapped}}$. Third, the selected configuration index $j$ is decomposed into an $n$-bit binary vector, forming the bit-flip mask $b_{\mathrm{flip}} \in {0,1}^{n \times \lceil M / row_len \rceil \times K}$, where $b_{\mathrm{flip}}[j,c,k] = 1$ indicates that slice $j$ of column $k$ in array $c$ has been flipped.

During inference, each partial sum (i.e. the CiM output) from a sub-array column with corresponding $b\_flip$ as 1 is adjusted by subtracting it from $\sum I$. $\sum I$ can be calculated using a shared adder tree (details in Section~\ref{sec-multibit:hardware-imp}). The required hardware additions are minimal: a near-memory register to store $b_{\mathrm{flip}}$, an adder tree (shared between multiple bit columns across multiple memory arrays to amortize its cost), a subtractor and a 2:1 MUX.  By enabling exponential growth in the search space and correcting slice-level polarity digitally, the bit-flip algorithm further improves tolerance to SAFs compared to both CVM and sign-flip. A detailed discussion of the effectiveness of bit-flip and its hardware implications is presented subsequently.

It is important to note that sign-flip and bit-flip are independent techniques for enhancing SAF tolerance. While their combination may offer additional benefits, exploring the feasibility of such a joint application is left for future work.

\subsection{Hardware Implementation of Sign-Flip and Bit-Flip}
\label{sec-multibit:hardware-imp}

\begin{figure}[t!]
\centering
  \includegraphics[width = 0.7\linewidth]{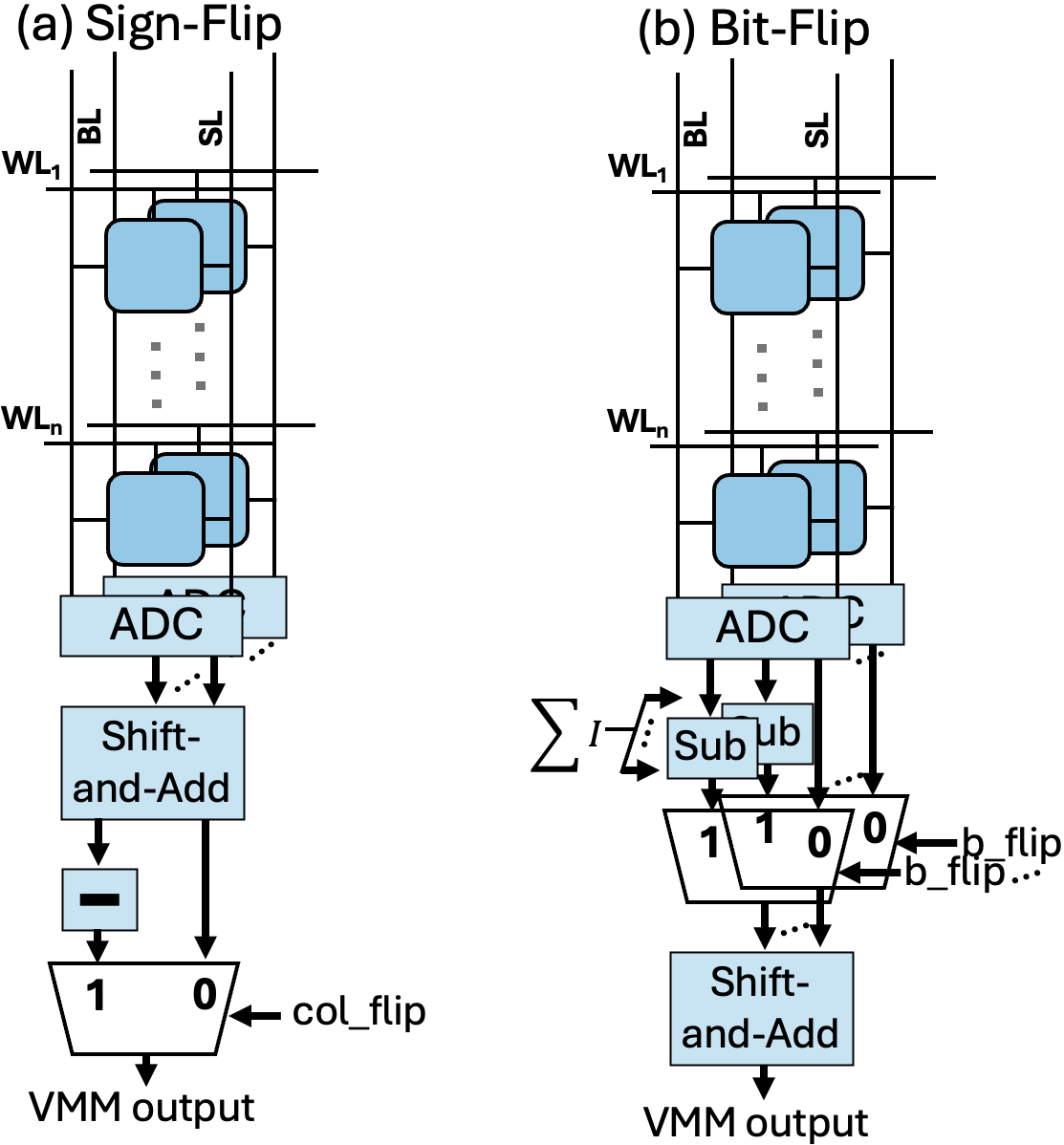}
  \caption{Crossbar sub-array columns and the peripheral circuits showing the hardware modifications required to enable (a) sign-flip and (b) bit-flip.}
  \label{fig-multibit:hardware}
\end{figure}

Although the proposed sign-flip and bit-flip algorithms are run in software (prior to weight deployment on on-chip memories), certain hardware modifications are needed to ensure functional correctness of matrix-vector multiplications during hardware operation (as noted before).   

Sign-flip requires a near-memory register to store $col\_flip$, a two's complement unit that multiplies the CiM output by $-1$, and a 2:1 multiplexer (MUX) to select between the original and negated CiM output. The two's complement unit can be implemented by first calculating the one's complement using inverters followed by adding one using a ripple carry adder. Since sign-flip is applied at the weight column level, one control bit per weight column is stored in $col\_flip$. Given that each weight column consists of multiple bit-columns (due to multi-bit weights), the hardware overhead is amortized across these bit-columns. Furthermore, because sign-flip modifies the final dot-product result, the multiplication by $-1$ is performed after the shift-and-add accumulation described in Section~\ref{sec-multibit:bit-slicing}. Thus, the hardware components required for sign-flip are shared across the bit-columns of a weight.

Bit-flip, in contrast, operates at the bit-column level, providing finer-grained control. It requires a near-memory register to store $b\_flip$, an adder tree to compute $\sum I$, a subtractor to compute $\sum I - \text{partial dot-product}$, and a 2:1 MUX to select between the original and flipped result. Because bit-flip operates at a finer granularity, the control bits and subtractors are required per bit-column. However, the adder tree can be shared across multiple bit-columns and even across memory arrays, as demonstrated in prior works \cite{nand_net}. Importantly, the post-processing in bit-flip occurs before the shift-and-add accumulation for each bit-column. While this finer granularity is expected to enable better SAF tolerance than sign-flip, it comes at the cost of larger hardware overhead (more details later).

Figure~\ref{fig-multibit:hardware} shows the hardware modifications required to enable sign-flip and bit-flip. In Section~\ref{sec-multibit:hardware_overhead} we quantify the energy, latency and area overheads of these modifications.
\section{Optimizing the Weight Transformations with LUT-based implementation}
\label{sec-multibit:lut}

While the expanded search space enabled by sign-flip and bit-flip techniques improves their ability to search and obtain SAF tolerant DNN weight mappings for bit-sliced crossbars, it also introduces challenges in terms of runtime and memory consumption (while running the algorithms in software). This reflects a fundamental trade-off: larger search spaces improve solution quality but are more expensive to explore.  

In sign-flip, the candidate tensor passed to the CVM routine has size $2 \times M \times K$. However, in bit-flip, this grows exponentially to $2^n \times M \times K$, where $M$ and $K$ denote the matrix dimensions and $n$ is the bitwidth. Although these tensors by themselves can be parallelized across GPUs for modest $n$ (e.g., $n \leq 8$), the bottleneck arises from the internal mechanics of the CVM algorithm itself (which we also utilize as a part of the bit-flip and sign-flip algorithms). Recall from Section~\ref{sec-multibit:closestvalue} that CVM performs an additional candidate enumeration over $2^n$ values for each incoming weight, effectively expanding the total search space to $2^n \times 2 \times M \times K$ for sign-flip and $2^n \times 2^n \times M \times K$ for bit-flip. For bit-flip in particular, this $2^{2n} \times M \times K$ scale becomes prohibitive in terms of memory usage, easily exceeding the available GPU memory, even on high performance GPUs like the NVIDIA A100.

A naive approach to mitigate this problem would be to process smaller chunks of the weight matrix (i.e. reducing $M$ and $K$) in more iterations. However, this slows down the mapping process significantly, defeating the goal of rapid deployment. To address this challenge, we propose a Lookup Table (LUT)-based optimization for CVM, which allows us to retain the benefits of the expanded search space without incurring any notable runtime or memory penalties. Our key insight is that CVM is a deterministic function, i.e. for any given input weight and SAF pattern, its output is fixed. Therefore, we precompute the output of CVM for all possible input combinations and store them in an LUT.

We construct a CVM lookup table (LUT) where each entry maps a unique pair of $(W_{\text{target}}, S)$ to its closest-value-mapped output $W_{\text{mapped}}$, with $W_{\text{target}} \in \{0, \dotsc, 2^n-1\}$ and $S \in \{-1, 0, 1\}^n$. The total number of possible inputs to this LUT is $2^n \times 3^n = 6^n$, since each of the $n$ bits can independently take one of three SAF states: fault-free (0), stuck-at-1 (+1), or stuck-at-0 (-1). This results in a LUT of size $6^n$, which remains tractable for $n \leq 8$. For example, with $n=8$, the LUT contains approximately 1.6 million entries, which easily fits within modern GPU memory.

With this precomputed LUT, both sign-flip and bit-flip algorithms can bypass the costly internal enumeration of CVM. Instead of invoking CVM dynamically, they perform a direct table lookup for each candidate weight, drastically reducing computational overhead. This effectively reduces the search space back to $2 \times M \times K$ for sign-flip and $2^n \times M \times K$ for bit-flip, which is easily manageable on typical GPUs.

Moreover, the LUT is model-agnostic and layer-agnostic, i.e. it needs to be constructed only once for a given bitwidth $n$ and can then be reused across all layers and models. This makes our approach a powerful and scalable optimization technique for any deployment pipeline that uses these fault-tolerant mapping strategies. We quantify the runtime improvements enabled by the LUT-based CVM implementation in Section~\ref{sec-multibit:speedup}.
\section{Results}
\label{sec-multibit:results}

\begin{figure*}[t]
    \centering
    \begin{subfigure}[b]{0.32\textwidth}
        \centering
        \includegraphics[width=\linewidth]{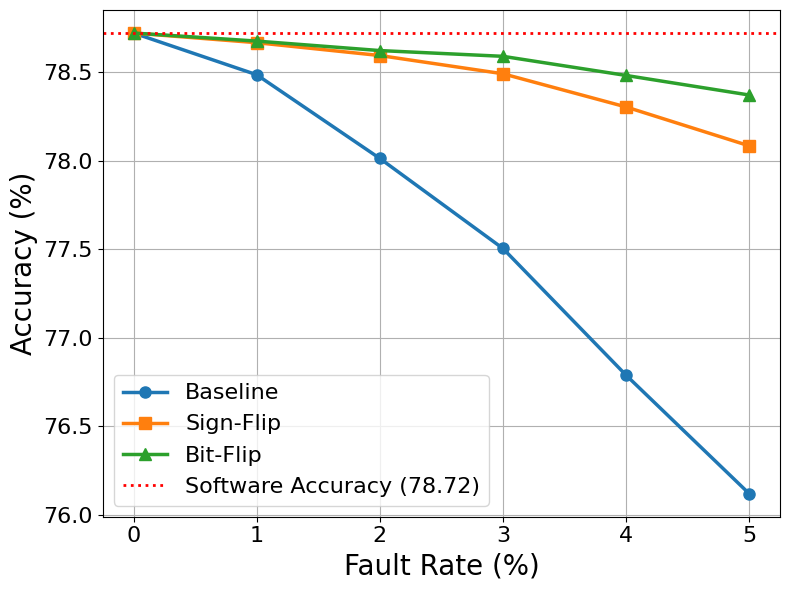}
        \caption{ResNet-18}
    \end{subfigure}
    \hfill
    \begin{subfigure}[b]{0.32\textwidth}
        \centering
        \includegraphics[width=\linewidth]{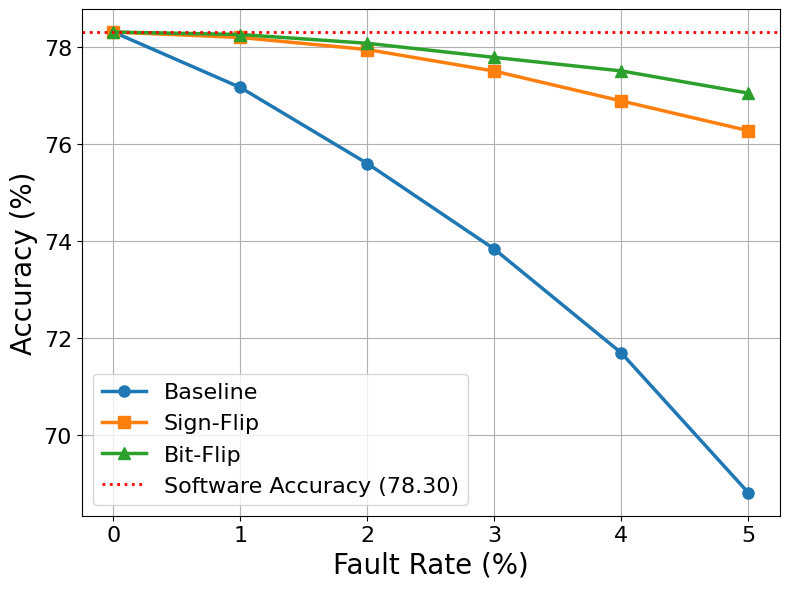}
        \caption{ResNet-50}
    \end{subfigure}
    \hfill
    \begin{subfigure}[b]{0.32\textwidth}
        \centering
        \includegraphics[width=\linewidth]{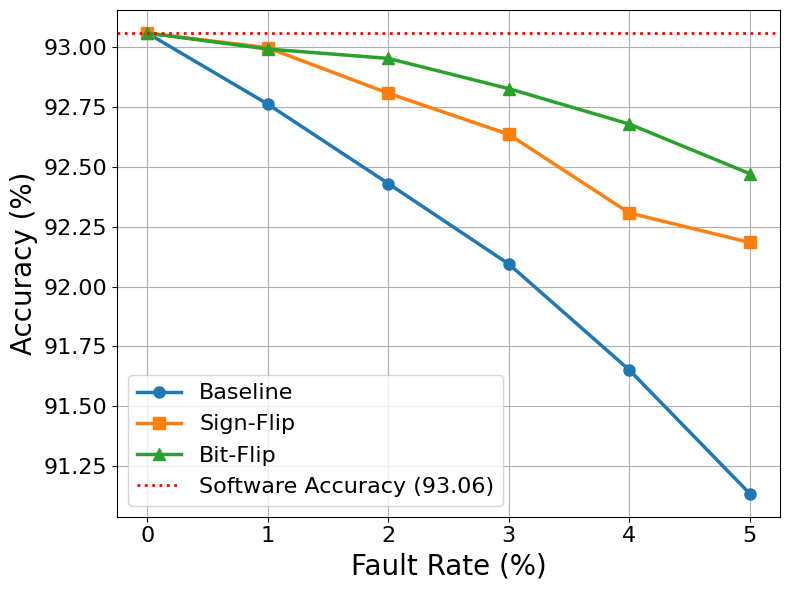}
        \caption{ViT}
    \end{subfigure}

    \caption{Performance results of various models on the CIFAR100 dataset. SAF rates up to 5\% are evaluated.}
    \label{fig-multibit:cifar100_results}
\end{figure*}

\begin{figure*}[t]
    \centering
    \begin{subfigure}[b]{0.32\textwidth}
        \centering
        \includegraphics[width=\linewidth]{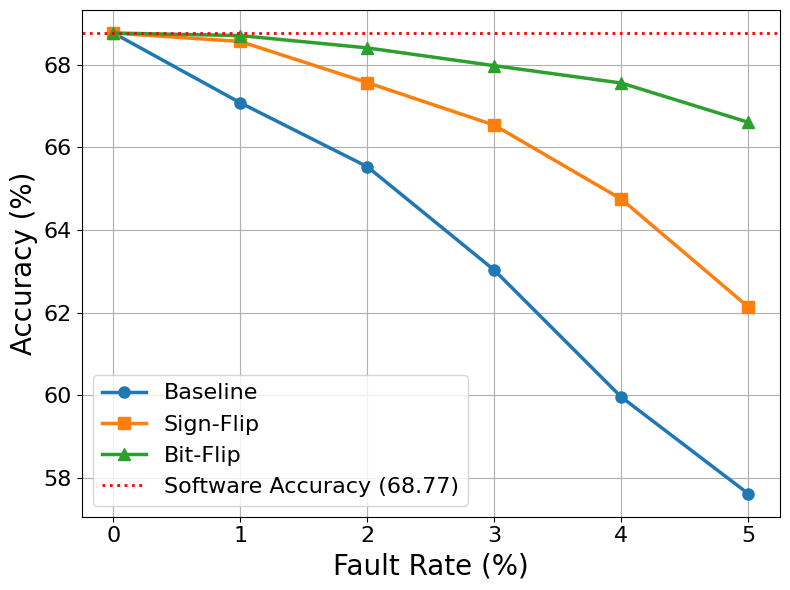}
        \caption{ResNet-18}
    \end{subfigure}
    \hfill
    \begin{subfigure}[b]{0.32\textwidth}
        \centering
        \includegraphics[width=\linewidth]{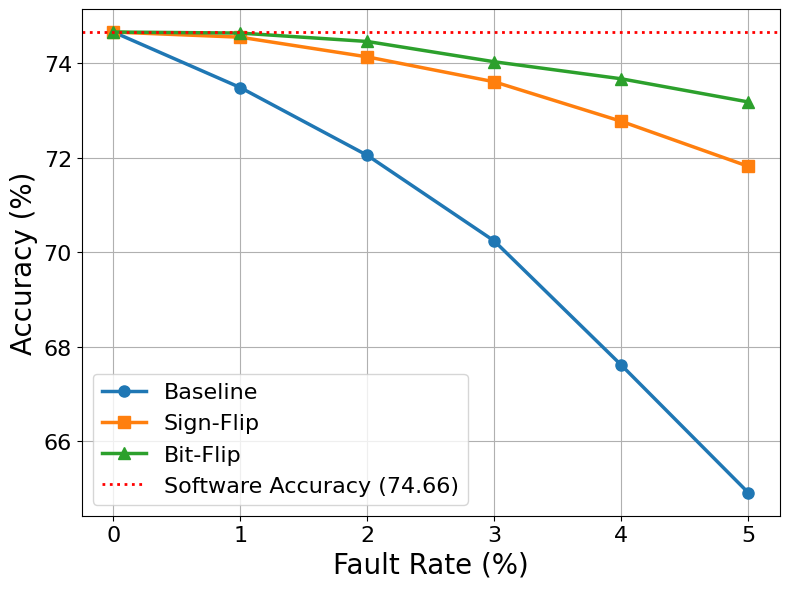}
        \caption{ResNet-50}
    \end{subfigure}
    \hfill
    \begin{subfigure}[b]{0.32\textwidth}
        \centering
        \includegraphics[width=\linewidth]{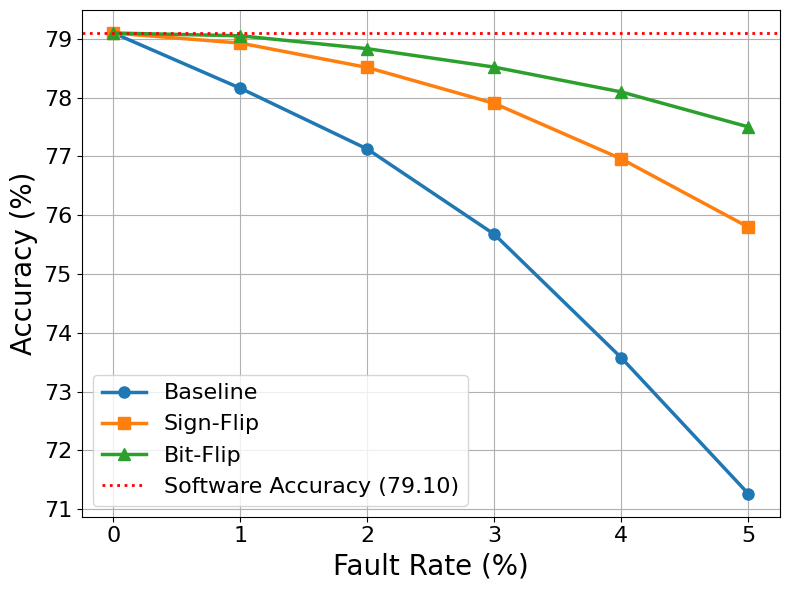}
        \caption{ViT}
    \end{subfigure}

    \caption{Performance results of various models on the ImageNet dataset. SAF rates up to 5\% are evaluated.}
    \label{fig-multibit:imagenet_results}
\end{figure*}

We evaluate the fault tolerance of three weight transformation techniques, CVM (baseline in this paper), sign-flip, and bit-flip, across fault rates ranging from 0\% to 5\% on several DNNs with 8-bit weights and activations. The evaluation spans diverse architectures and datasets, including ResNet-18 and ResNet-50 on CIFAR-100 and ImageNet, as well as Vision Transformers (ViT) on CIFAR-100 and ImageNet. We obtain the pretrained weights for the models from Hugging Face for CIFAR-100 models and torchvision for the ImageNet models\cite{resnet18-cifar100,resnet50-cifar100,vit-cifar100,torchvision}. We utilize the post-training quantization scheme in \cite{google_quant} to quantize the weights and activations to 8-bit, observing negligible loss in accuracy. We simulate these models on CiM hardware designed with 64x64 memory sub-arrays. 

For our fault injection experiments, we inject SAFs (both SA1 and SA0) randomly and uniformly in the memory sub-arrays across the layers and weights of the DNNs. We perform 50 Monte Carlo simulations for each fault rate, and present the average inference accuracy.

In the ResNet models, the convolutional layers involve fixed weights and can be expressed as matrix multiplications, making them well-suited for mapping onto CiM-enabled memory arrays. In contrast, Vision Transformers (ViTs) involve two distinct types of matrix multiplications during inference: (i) in the feedforward layers, and (ii) in the self-attention layers. In the feedforward layers, multiplication occurs between static weight matrices and dynamic input activations. However, the self-attention layers perform multiplications between query, key, and value matrices that are generated on-the-fly from the input representations. Because non-volatile memory technologies like ReRAM and FeFET suffer from high write latency, energy overhead, and limited endurance, they are poorly suited for storing or frequently updating such dynamic matrices. As a result, prior works such as \cite{heterogenous, all_digital} offload self-attention computations to digital cores while reserving CiM arrays for the feedforward layers. We adopt the same strategy in our accelerator, treating self-attention layers as SAF-free and assuming that stuck-at faults (SAFs) occur only in the feedforward weights.

\subsection{SAF Tolerance}

Fig.~\ref{fig-multibit:cifar100_results} and Fig.~\ref{fig-multibit:imagenet_results} show the accuracy results from the fault injection experiments. Across all models and datasets, we observe that increasing SAF rates lead to a consistent decline in accuracy for the baseline CVM approach, with ImageNet experiencing larger drops than CIFAR-100. However, when the proposed sign-flip/bit-flip is applied along with CVM, it significantly reduces this degradation by introducing flexibility in weight mapping and mitigating the impact of faults without retraining. 

On ResNet-18 and ResNet-50, the baseline shows substantial degradation of up to $\sim$11\% (at 5\% fault rate). Sign-flip reduces this degradation by about half, while bit-flip consistently achieves even greater resilience, typically bringing the final accuracy within 1-2\% of the ideal accuracy.

For ViT on CIFAR-100, the accuracy drop is modest, up to 2\%. Both sign-flip and bit-flip reduce it further, with bit-flip maintaining accuracy within 1\% of the ideal. On ImageNet, however, the baseline suffers a more pronounced drop ($\sim$8\%), while sign-flip cuts this down to approximately 3\%. bit-flip goes further, limiting the degradation to only 1.6 percentage points. 

In summary, both proposed techniques outperform the baseline in all tested scenarios. Sign-flip provides a simple and effective method for reducing SAF-induced degradation, while bit-flip offers more fine-grained control for weight mapping and brings model accuracy significantly closer to the ideal, across the entire range of fault rates considered. These results demonstrate the utility of our training-free weight transformations for fault-aware deployment of DNNs on bit-sliced crossbars.

\subsection{Algorithm Runtime Analysis}
\label{sec-multibit:speedup}

\begin{table}[t]
\centering
\caption{Runtime (s) of fault injection and mitigation with and without LUT optimization}
\label{tab:execution_time}
\resizebox{\columnwidth}{!}{%
\begin{tabular}{lccc}
\toprule
\textbf{Model} & \textbf{Baseline} & \textbf{Sign-Flip} & \textbf{Bit-Flip} \\
 & (LUT / no-LUT) & (LUT / no-LUT) & (LUT / no-LUT) \\
\midrule
ResNet-18 & 0 / 10   & 2 / 21    & 16 / 1200  \\
ResNet-50 & 0 / 22   & 3 / 49    & 30 / 2640  \\
ViT       & 1 / 95   & 4 / 200   & 120 / 10800 \\
\bottomrule
\end{tabular}%
}
\end{table}

In Section~\ref{sec-multibit:lut}, we introduced the LUT-based CVM approach, which is used to accelerate our fault mitigation algorithms. Recall that the naive exploration (without the LUT optimization) of the entire search space enabled by sign-flip and bit-flip is very time- and memory-intensive. Here, we quantify the speedup resulting from the proposed LUT-based approach by measuring the runtime of these algorithms on an NVIDIA T4 GPU across various models, both with and without LUT-based optimization. As shown in Table~\ref{tab:execution_time}, the LUT-based approach yields substantial reductions in execution time across all techniques and architectures.

For instance, bit-flip-based solution search on ViT takes over 3 hours (10,800 seconds) without LUTs, but only 2 minutes (120 seconds) with them. Similarly, sign-flip-based search on ResNet-50 reduces from 49 seconds to just 3 seconds. Even CVM (baseline) sees runtime drop from 10-95 seconds down to less than 1 second.

This level of acceleration enables our framework to support the fast computation of effective fault-tolerant mappings. As a result, we can efficiently assess robustness and restore accuracy to near-ideal levels, making the framework highly suitable for large-scale deployment studies and design-space exploration.

\subsection{Hardware Overhead}
\label{sec-multibit:hardware_overhead}


\newcolumntype{Y}{>{\raggedright\arraybackslash}X}

\begin{table}[t]
\centering
\caption{Hardware overhead (\%) of Sign-Flip (SF) and Bit-Flip (BF) for different memory technologies}
\label{multibit:table:overhead}
\small
\setlength{\tabcolsep}{4pt}
\renewcommand{\arraystretch}{1.2}
\begin{tabularx}{\columnwidth}{Y c c c c}
\toprule
\textbf{Tech} & \textbf{Method} & \textbf{Energy} & \textbf{Latency} & \textbf{Area} \\
\midrule
SRAM  & Sign-Flip & 0.2 & 1.0 & 1.3 \\
SRAM  & Bit-Flip & 4.1 & 2.8 & 3.8 \\
ReRAM & Sign-Flip & 0.3 & 1.4 & 1.6 \\
ReRAM & Bit-Flip & 4.6 & 3.7 & 4.4 \\
FeFET & Sign-Flip & 0.3 & 1.3 & 1.6 \\
FeFET & Bit-Flip & 4.6 & 3.6 & 4.5 \\
\bottomrule
\end{tabularx}
\end{table}

We evaluate the hardware implications of sign-flip and bit-flip at the memory macro level, which includes the crossbar arrays and their associated peripheral circuitry. While our analysis is limited to the memory macro, it is important to note that the system-level overheads may be even lower when accounting for additional components beyond the memory subsystem.

The operation under evaluation is a single in-memory vector-matrix multiplication. Specifically, an 8-bit input vector of length 64 is processed by eight  $64 \times 64$ crossbar sub-arrays storing 8-bit weights to produce 64 dot products in one cycle. We design the crossbar arrays using three memory technologies: 8T-SRAM, 1T-1ReRAM, and 1FeFET. All arrays employ current-sensing techniques to generate their outputs. For the SRAM arrays, we use the 7nm Predictive Technology Model (PTM) \cite{ptm}. For ReRAM, we adopt an experimentally calibrated compact model for Al-doped $HfO_X$ devices from \cite{rram}. The FeFET arrays are modeled using a compact model of $Hf_{0.5}Zr_{0.5}O_{2}$-based FeFETs, in which the ferroelectric behavior is captured using modified Preisach equations. The ferroelectric capacitor is coupled with a 7nm transistor \cite{fefet_model}. This model is calibrated with experimental results reported in~\cite{fefet_experiments}. More details of these models can be found in \cite{bitcells}.

To estimate the energy, latency and area of each crossbar sub-array, we first develop custom layouts based on the methodology described in~\cite{bitcells}. Once the physical dimensions of the arrays are obtained in terms of gate, metal, and fin pitches, we estimate parasitic wire resistances and capacitances using models from~\cite{para_res, para_cap}. These parasitics are then incorporated into SPICE simulations, which are used to determine the overall energy, latency, and physical area (derived from the layout dimensions) for each array.

To estimate the energy, latency, and area of the memory array peripherals, including the flash ADC, shift-and-add circuitry, as well as the additional peripherals required by sign-flip and bit-flip, we use NeuroSim~\cite{neurosim}. The estimates are also based on the 7nm Predictive Technology Model (PTM)~\cite{ptm} for both the peripheral circuits and post-processing units.

To keep array non-idealities such as the effect of parasitic resistances and device non-linearities in check, we utilize partial word-line activation (PWA), simultaneously activating 16 rows (out of 64) at a time. PWA also reduces the required ADC precision, enabling us to use 4-bit flash ADCs, albeit at the cost of latency. To further minimize ADC area overhead, we share one set of column peripherals (ADCs + post-processing circuitry) across eight columns, again trading off latency for improved area efficiency. The adder tree in bit-flip is also shared across the 8 crossbar arrays since they have common activation vectors. 

Table~\ref{multibit:table:overhead} presents the energy, latency, and area overheads incurred by sign-flip and bit-flip, relative to the baseline that uses closest value mapping (CVM) only. Across all three memory technologies, the overheads are modest. Bit-flip consistently incurs higher overheads than sign-flip, which is attributed to its finer granularity of operation and the need for more hardware. Further, while sign-flip operates at the weight-column level, bit-flip operates at the bit-column level, increasing the amount of additional processing required. For example, in bit-flip, each CiM output corresponding to a bit slice must be individually subtracted with input sum. This subtraction must be repeated for \textit{every} bit slice. In contrast, sign-flip performs a single negation on the final dot product after the bit-sliced outputs have been shifted and accumulated. As a result, sign-flip processes fewer values, resulting in lower overhead.

Quantitatively, with respect to the baseline, the energy overheads for sign-flip range between 0.2\% and 0.3\%, while those for bit-flip range between 4.1\% and 4.6\%. The latency overheads follow a similar pattern: sign-flip adds only 1.0\% to 1.4\%, while bit-flip adds 2.8\% to 3.7\%. Area overheads arise mainly from storing the column-level control vectors-$col\_flip$ for sign-flip and $b\_flip$ for bit-flip. Since sign-flip requires only one additional bit per weight column (shared across 8 bit-columns), its area overhead is lower, ranging from 1.3\% to 1.6\%. Bit-flip requires one bit per bit-column, resulting in area overheads ranging from 3.8\% to 4.5\%.

It is important to emphasize that the total energy, latency, and area of the memory macro are largely dominated by the ADCs and the memory arrays. As a result, the peripheral overheads introduced by sign-flip and bit-flip remain relatively insignificant. With mindful peripheral design, both techniques can be integrated efficiently and provide substantial resilience to SAFs. While bit-flip offers superior fault tolerance compared to sign-flip, it does so at a higher hardware cost. Sign-flip, on the other hand, provides meaningful error mitigation with lower overhead, making it a compelling option in highly resource-constrained environments. In scenarios where additional hardware overhead is acceptable, bit-flip may be preferred to achieve higher robustness.
\section{Conclusion}
\label{sec-multibit:conclusion}

In this paper, we proposed two novel, training-free weight transformation techniques, sign-flip and bit-flip, to improve the SAF tolerance of DNNs with multi-bit weights and activations,  deployed on bit-sliced compute-in-memory (CiM) accelerators. These techniques enhance SAF tolerance by expanding the search space for SAF-aware weight mappings. Sign-flip operates at the weight-column level by choosing between a weight and its negation. On the other hand, bit-flip provides fine-grained control by selectively inverting individual bit slices. By introducing polarity as an additional axis of optimization, both techniques work in conjunction with closest value mapping (CVM) to significantly improve the ability to map weights onto faulty memory arrays without any training required. 

To enable efficient deployment, we introduced a LUT-based implementation of CVM that drastically reduces the runtime and memory demands of both techniques, making them scalable to large DNN models. We also developed a comprehensive framework for SAF-mitigation strategies, allowing for rapid exploration across models and SAF rates.

Extensive experiments on ResNet-18, ResNet-50, and ViT across CIFAR-100 and ImageNet show that sign-flip and bit-flip consistently reduce SAF-induced accuracy degradation. Bit-flip offers the highest fault resilience, often recovering accuracy to within 1-2\% of the ideal baseline, while sign-flip delivers meaningful improvements at lower hardware cost. Our hardware analysis for SRAM, ReRAM, and FeFET technologies shows that both techniques can be integrated with minimal energy/latency/area overhead . These results highlight the practicality and scalability of sign-flip and bit-flip as effective fault-mitigation strategies for large-scale CiM systems.
\section{Acknowledgements}
\label{sec:acknowledgements}

This work is supported by the Center for the Co-Design of Cognitive Systems (COCOSYS), one of seven centers in JUMP 2.0, funded by Semiconductor Research Corporation (SRC) and DARPA, and by Raytheon and NSF.

\bibliography{references.bib}
\bibliographystyle{IEEEtran}

\end{document}